# Stagnation of electron flow by a nonlinearly generated whistler wave

Toshihiro Taguchi[1]

[1] Department of Electrical and Electronic Engineering, Setsunan University, Neyagawa, Osaka, Japan

Thomas M. Antonsen Jr.[2]

[2] Institute for Research in Electronics and Applied Physics, University of Maryland, College Park, MD, U.S

Kunioki Mima[3]

[3] The Graduate School for the Creation of New Photonics Industries, Hamamatsu, Shizuoka, Japan

## Abstract

Relativistic electron beam transport through a high-density, magnetized plasma is studied numerically and theoretically. An electron beam injected into a cold plasma excites Weibel and two-stream instabilities that heat the beam and saturate. In the absence of an applied magnetic field, the heated beam continues to propagate. However, when a magnetic field of particular strength is applied along the direction of beam propagation, a secondary instability of off-angle whistler modes is excited. These modes then couple nonlinearly creating a large amplitude parallel propagating whistler that stops the beam. In this letter, we will show the phenomena in detail and explain the mechanism of whistler mediated beam stagnation.

Transport and scattering of beams of energetic electrons in magnetized plasma are long studied problems with applications in space physics, plasma astrophysics and laboratory plasma physics [1-3]. They are particularly important in the fast ignition inertial confinement fusion scheme [4]. Here a pellet of compressed D-T fuel is ignited by a beam of energetic electrons, created at a distance by an ultra-short, ultra-intense laser pulse. To reach the pellet, the beam must propagate through dense plasma from where it is created, the critical surface for the laser pulse, to the location of the compressed pellet.

Typically, beam plasma interactions are considered from the starting point of linear theory. An electromagnetic mode is found to grow linearly, extracting energy from the beam, until the distribution of electron momenta is altered and growth of the mode is arrested. The beam is not necessarily stopped, rather it continues to propagate, albeit with a broadened momentum distribution. In this letter we present an example of a nonlinear instability in which a large amplitude whistler wave, which is co-propagating with the beam and is linearly stable, is nonlinearly pumped by obliquely propagating whistler waves, and grows until the beam is stopped.

As mentioned, a motivation for our study is the fast ignition scheme, the advantage of which is that by using the beam generated by the short pulse laser, the energy demands on the compressing laser are significantly reduced [4]. An issue with the fast ignition scheme is that the electron beam, on its way to the pellet, scatters off self-generated fluctuations, reducing the energy flux density on the pellet. One set of fluctuations that appears is associated with the electrostatic two-stream instability, which saturates mainly by heating the parallel momentum components of the beam. A second set of fluctuations that appears is associated with the Weibel instability [5]. Specifically, the high-energy electron beam quickly induces a return current in the cold plasma, and the counter-streaming electrons are then unstable to the generation of transverse magnetic fields with transverse structure. This instability saturates when the transverse temperature of the beam reaches a sufficient level. However, at this level the beam is not well-collimated.

In order to suppress electron deflection, several methods have been proposed. One such method is the application of a strong external magnetic field along the direction of beam propagation [6]. In the experiments of Ref. [6], a kilo-tesla magnetic field is generated by a second intense laser-plasma interaction. When such a strong magnetic field is applied, it is expected that the electron beam will remain collimated. In addition, the growth of the Weibel instability should also be reduced as the magnetic field restricts the transverse electron motion. This latter effect is confirmed in both our theory and simulations. However, we find the surprising result that within a broad range of magnetic fields the beam is stopped by the appearance of a large amplitude whistler wave.

We conducted simulations using a 2 1/2-dimensional hybrid code in which the energetic electrons are treated as particles, while the background electrons are treated as a fluid. The ions are assumed to be immobile. Results of sample simulations are shown in Figs. 1 and 2. The simulation domain is $0< z/d_e < 750$ in the direction of beam propagation ($z$) and $0 < x/d_e< 125$ in one direction transverse to the direction of beam propagation ($x$), where $d_e$ is the collisionless skin depth $d_e = c / \omega_{pe}$, and the plasma frequency $\omega_{pe} = \sqrt{e^2 n_0 / \varepsilon_0 m_e}$ based on the background electron density $n_0$ and the rest mass of an electron $m_e$. The boundary conditions in $x$ are that all quantities are periodic. The boundary conditions in $z$ are more complicated. Two layers are added to the ends of the simulations to damp any waves and absorb any electrons impinging on the end boundaries. An energetic beam of electrons is introduced in the layer near $z$=0, the boundary on the left, with a distribution in momentum given by, $f(\boldsymbol{p}) = A\exp\left[\left(m_e\gamma c^2 - \boldsymbol{p}\bullet\boldsymbol{u}_0\right)/T_b\right]$. Here, $\boldsymbol{p}$ is the momentum of each particle, $\gamma = \sqrt{1+\boldsymbol{p}^2/m_e^2c^2}$ is its Lorentz factor, $\boldsymbol{u}_0$ is the drift velocity and $T_b$ is the temperature of the beam electrons.

Parameters in the simulation are set as follows: beam velocity, $u_0$= 0.95$c$, injected beam density, $n_b=n_0/10$, beam temperature, $T_b$=100keV, while the background electron temperature $T_e$ is 10keV. A typical background electron density is $n_0=10^{22}$ cm$^{-3}$, which corresponds to about 10 times of the critical density for a 1μ–laser pulse.

The bottom image of Fig.1 displays the beam density (color scale indicates density relative to the background density) at $\omega_{pe}t$ =1600 with no applied magnetic field. This is well after the front of the beam has passed through the simulation domain and a statistically steady state has formed. The injected beam distribution is Weibel unstable and the beam forms filaments and heats in the transverse direction. This is also illustrated by the line plots attached in Fig.1. These are line plots of the following $x$-averaged quantities: the normalized beam density ($\bar{n}_h$), the normalized in-plane ($\bar{B}_x$) and out-of-plane ($\bar{B}_y$) magnetic field fluctuations, and the longitudinal electric field fluctuations ($\bar{E}_z$). Here the magnetic fields are normalized by $m_e\omega_{pe}/e$ and the electric field is normalized by $m_e\omega_{pe}c/e$. Since the beam heats as it propagates, the Weibel instability saturates and the magnetic field fluctuations decay. Thus, the heated beam crosses the entire simulation domain as evidenced by the line plot of the beam density. At the time of this image the ratio of beam density leaving the right boundary to that injected is 1.25. This small density increase is caused by the decrease of the flow velocity due to beam electron scattering by the Weibel instability.

The situation is different when a strong magnetic field is applied along the direction of beam propagation. This is illustrated in Fig. 2 for the case of the normalized external B-field, $\omega_c / \omega_{pe} = 0.3$, where $\omega_c = eB_0/m_e$ for the external magnetic field $B_0$. This value corresponds to 9.6 kT in the case of $n_0=10^{22}$ cm$^{-3}$. In the simulation, the Weibel instability grows and heats the beam, as evidenced by the out-of-plane magnetic field fluctuations. However, after the beam propagates to a distance $z/d_e \sim 200$, both in-plane and out-of-plane magnetic fields grow and the beam is reflected. This is further illustrated by the line plot of the beam density, where the down stream density is half that of Fig. 1.

The in-plane and out-of-plane components of the magnetic field for $180 < z/d_e < 300$ in Fig. 2 have the characteristics of a circularly polarized wave: they vary sinusoidaly in $z$ with wavelength approximately $\lambda \approx 16\ d_e$, and they are 90 degrees out of phase. We thus identify the disturbance as a whistler wave. The dispersion relation for long-wavelength ($\lambda >> d_e$) low frequency ($\omega << \omega_c$) whistler waves propagating at an angle to the magnetic field in cold plasma is [7]

$$\omega = \frac{\omega_c c^2}{\omega_{pe}^2} k k_z \,, \qquad (1)$$

where $k = \sqrt{k_x^2 + k_z^2}$. Here $\omega$ is the wave frequency and $\boldsymbol{k} = k_x \hat{\boldsymbol{x}} + k_z \hat{\boldsymbol{z}}$ is the wave vector. For parallel propagation ($k_x$ =0), as applies to the $x$-averaged fields, the dispersion relation implies a positive phase velocity and a positive group velocity along the magnetic field. The group velocity increases with magnetic field strength, the implications of which will be discussed subsequently.

Direct excitation of the purely parallel-propagating whistler by wave particle energy exchange involving the energetic stream of electrons is not expected. Wave particle interaction is possible when the resonance condition

$$\gamma\omega - n\omega_c - k_z p_z / m_e = 0 \,, \tag{2}$$

is satisfied, where the integer $n = 1, 0, -1$ denotes cyclotron, Cherenkov, and anomalous cyclotron resonance respectively [7]. These resonances are mediated by different components of the electromagnetic field. The Cherenkov resonance is mediated by the axial electric field (which is small for whistlers), and the cyclotron and anomalous resonances are mediated by one or the other circular polarizations of the transverse electric field. The whistler wave is circularly polarized in the same sense of rotation as is the gyro-motion of electrons, which means that only the $n$ =1 resonance is active for purely parallel propagation. Thus, for the low frequency whistler wave ($\omega << \omega_c$) only counter propagating electrons can be resonant, and since the wave phase velocity is positive, energy will flow from particles to fields only if the perpendicular temperature of these counter-propagating electrons is greater than parallel temperature. This is the mechanism of the heat flux instability studied extensively by Gary [8]. To tap the energy of the forward propagating hot electrons it is necessary to consider off-angle propagation, for which the transverse wave electric field becomes elliptically polarized and nonzero Larmor radius effects enter. Both of these effects activate the $n$ = -1 resonance and energy can be transferred from electrons to fields as the electrons lower their energy while increasing their perpendicular momentum in the presence of a wave with a positive phase velocity [7].

In our simulations the mechanism of excitation of the parallel propagating whistler wave, once the Weibel has stabilized, is by the growth of hot electron driven obliquely propagating whistlers excited through the anomalous Doppler resonance. The oblique whistlers then nonlinearly couple to the weakly damped parallel propagating whistler, which grows with a wavelength determined by the condition that its group velocity is sufficiently small so that the wave remains close to the point of injection of the electron beam. This situation arises specifically in our simulations due to the fact that the beam is continually injected from one end.

In order to investigate the beam dynamics theoretically we have performed a linear stability analysis based on solution of the linearized, relativistic Vlasov equation coupled with Maxwell's equations [9]. This system leads to a 3 by 3 tensor equation for the components of the electric field. The vanishing of the determinant of the tensor gives rise to a dispersion relation,

$$\left| c^2 k^2 - \omega^2 \mathrm{D} - c^2 \boldsymbol{kk} \right| = 0 \tag{3}$$

In forming the tensor D we assume that there are two populations of electrons (hot and cold) having different drift velocities and temperatures. To help in the evaluation of the elements of the tensor D we use a simplified relativistic Maxwell's distribution [6],

$$f_{0s}(\boldsymbol{p}) = \frac{n_s}{\left(2\pi m_e \gamma_s^{5/3} T_s\right)^{3/2}} \exp\left[ -\frac{p_x^2 + p_y^2}{2m_e \gamma_s T_s} - \frac{(p_z - p_s)^2}{2m_e \gamma_s^3 T_s} \right] \tag{4}$$

Here, the quantities $\gamma_s T_s$ and $\gamma_s^3 T_s$ represent perpendicular and parallel temperatures of specie-s, and $n_s$ and $p_s$ are the specie density and mean parallel momentum. The advantage of using this distribution is that the elements of D can be approximated using the familiar nonrelativistic plasma dispersion function, and using modified Bessel functions to account for nonzero Larmor radius effects. Details of the evaluations may be found in Refs. 10 and 11.

Solving Eq. (3) numerically, we obtain the growth rate, Im[$\omega$], of the most unstable mode as a function of wave vector ($k_z, k_x$). These results are plotted as false color images in Fig 3 for the case of (a) $\omega_c / \omega_{pe}$=0.01, (b) $\omega_c / \omega_{pe}$=0.3, and (c) $\omega_c / \omega_{pe}$ =0.7. In the three cases the parameters of the electron beam, such as the drift velocity, temperature and density, are the same as those in Figs. 1 and 2. The drift velocity and the number density of the background electrons are determined by the charge and current neutrality conditions. We then pick the parameters for the simplified distribution, Eq. (4), to populate a range of parallel momentum similar to what is observed in the simulation, specifically we use $\gamma_b$ =2, which corresponds to $p_s$=1.9$m_e c$.

As shown in Fig. 3 there are unstable modes occupying different regions of wavenumber space. One mode occupies a region of wavenumber space with growth rate maximum for $ck_z / \omega_{pe} \approx 1.3$ and $k_x = 0$, and is identified as a two-stream instability. The phase velocity of the mode with peak growth rate is found to be $\omega / k_z = 0.7c$. Thus, the mode is in Cherenkov resonance with the hot electrons. This mode is suppressed but not eliminated by the applied magnetic field. Instability also appears in the region where $ck_z < \omega_{pe}$ and $k_x c \approx \omega_{pe}$. Two different mode types are present depending on the strength of the magnetic field. For low magnetic field values as in Fig 3a growth is peaked at $k_z = 0$ and the real part of the mode frequency is small. We identify this mode as the Weibel instability. For larger magnetic field values as indicated in Fig. 3b the Weibel instability is suppressed. However, growth occurs for values of wavenumber $ck_z / \omega_{pe} \approx 0.1$ and $ck_x / \omega_{pe} = 0.5$. A plot of the real frequency and growth rate of this mode is shown in Fig. 3d and indicates that the mode is an off angle whistler as described by Eq. (1).

In order to analyze the dynamics of unstable modes in the simulation, we have performed a spatial two-dimensional Fourier transform of the out-of-plane magnetic field $\bar{B}_y$ in the region near the simulation boundary where electrons are injected (50 < $z/d_e$ < 175) (not shown). These plots confirm the sequential growth and saturation of the two-stream and Weibel spectral components. Appearing later in the simulation, at the time of stagnation, are modes with wavenumber $ck_z / \omega_{pe} \sim 0.1$. We believe these modes are the obliquely propagating whistler modes, possibly seeded by the nonlinear coupling of the two-stream modes. The initially unstable two-stream modes with short wavelength saturate and disappear. Thus, by the time $\omega_{pe} t$ =500 the oblique whistler modes dominate. The spectra show by the end of the simulation that the oblique modes are condensed into a single whistler wave with a large amplitude in agreement with Fig. 2.

In conclusion, we have found that a beam of energetic electrons in a strong magnetic field can drive up a large amplitude, *linearly stable* whistler wave through the excitation and nonlinear coupling of obliquely propagating whistler waves. The parallel propagating wave grows until the electron beam is reflected. This mechanism is likely to be important to efforts to collimate hot electrons in fast ignition fusion experiments. It may also be important in a wide range of space and astrophysical plasma settings.

This work was supported in part by JSPS KAKENHI Grant Number 15H03758, the Japan/U.S. Cooperation in Fusion Research and Development, and by the U.S. Department of Energy, Office of Science, Office of Fusion Energy Science, under Award Number DEFG0293ER54197. The authors acknowledge discussions with J. Drake, M. Swisdak, and G. Roberg-Clark.

## Figure Captions

Figure 1. The spatial distribution of the beam electron density at $\omega_{pe}t = 1600$ (bottom) and line plots of field quantities averaged over transverse direction (top) in the case of no external magnetic field.

Figure 2. The spatial distribution of the beam electron density at $\omega_{pe}t = 1600$ (bottom) and the line plots of field quantities averaged over transverse direction (top) in the case that an external magnetic field ($\omega_c / \omega_{pe}$ =0.3) is applied.

Figure 3: Growth rates of theoretically predicted unstable modes in two dimensional wavenumber space for different external magnetic fields, $\omega_c / \omega_{pe}$ =a) .01, b) .30 and c) .70, and d) growth rate and frequency of oblique whistler vs $k_z$ for selected values of $k_x$ for parameters of case c).

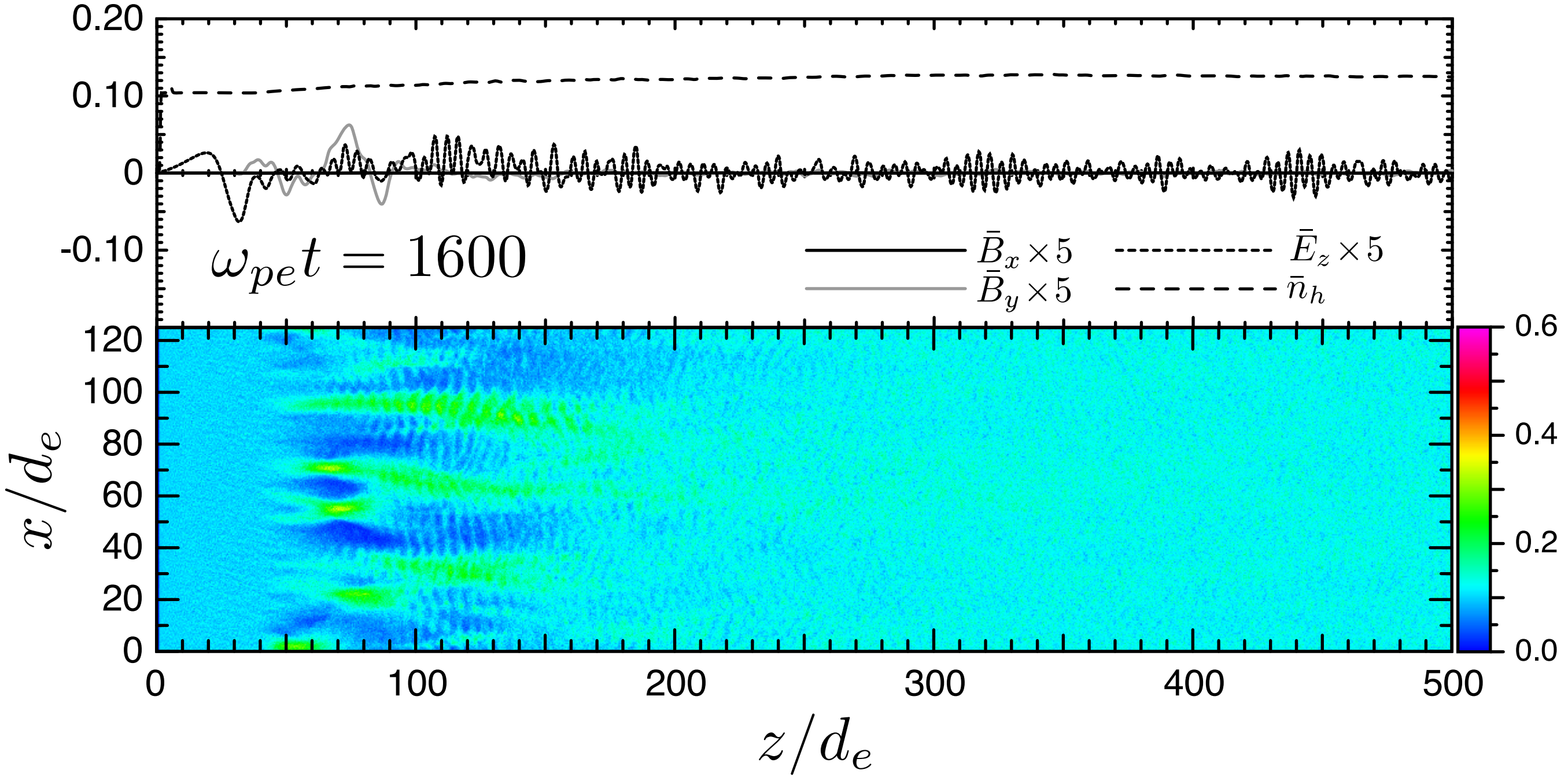

Figure 1

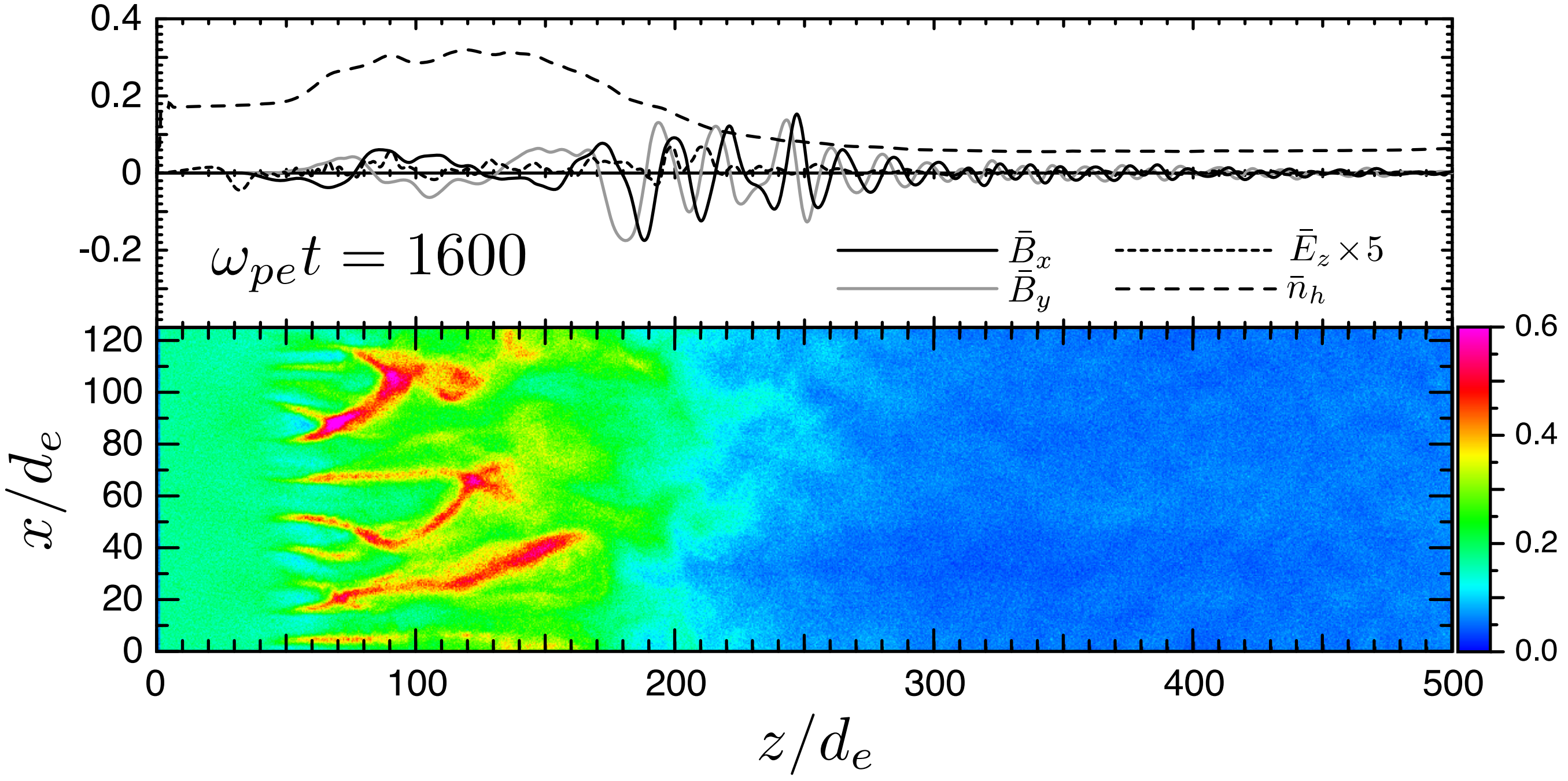

Figure 2

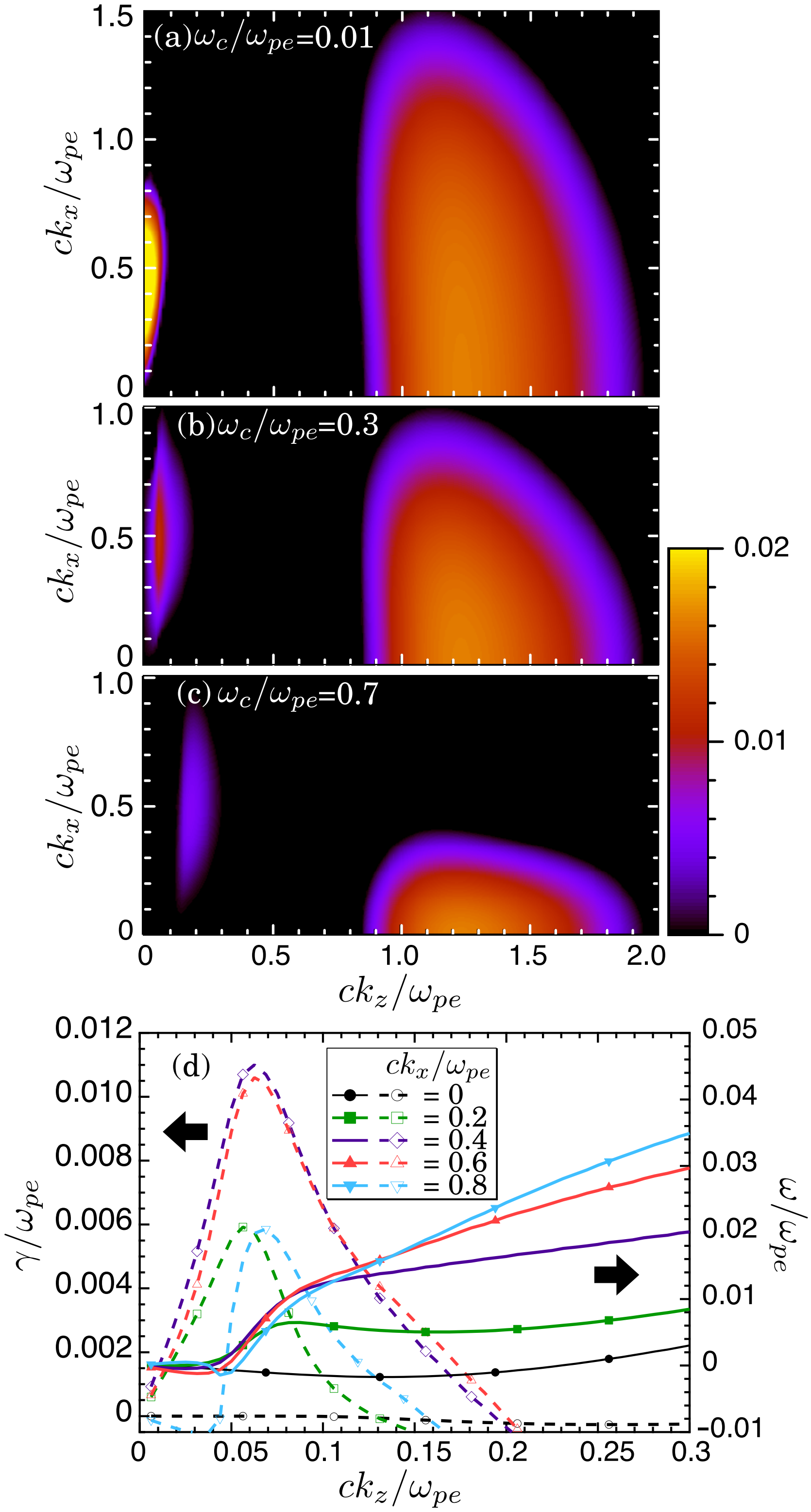

(a) $\omega_c/\omega_{pe}$=0.01
(b) $\omega_c/\omega_{pe}$=0.3
(c) $\omega_c/\omega_{pe}$=0.7
$ck_x/\omega_{pe}$
$ck_z/\omega_{pe}$
0 0.5 1.0 1.5 2.0
0 0.01 0.02
(d)
$ck_x/\omega_{pe}$
= 0
= 0.2
= 0.4
= 0.6
= 0.8
$\gamma/\omega_{pe}$
$\omega/\omega_{pe}$
0 0.05 0.1 0.15 0.2 0.25 0.3
0 0.002 0.004 0.006 0.008 0.010 0.012
-0.01 0 0.01 0.02 0.03 0.04 0.05

Figure 3